# Empirical Analysis on Productivity Prediction and Locality for Use Case Points Method

**Mohammad Azzeh [1]\*, Ali Bou Nassif [2], Cuauhtémoc López Martín[3]**

1   Department of Software Engineering, Applied Science Private University, Jordan; m.y.azzeh@asu.edu.jo
2   Department of Computer Engineering, University of Sharjah, UAE, anassif@sharjah.ac.ae
3   Department of Information Systems, Universidad de Guadalajara, Zapopan, Jalisco, México, cuauhtemoc@cucea.udg.mx

**Abstract.** Use Case Points (UCP) method has been around for over two decades. Although, there was a substantial criticism concerning the algebraic construction and factors assessment of UCP, it remains an efficient early size estimation method. Predicting software effort from UCP is still an ever-present challenge. The earlier version of UCP method suggested using productivity as a cost driver, where fixed or a few pre-defined productivity ratios have been widely agreed. While this approach was successful when no enough historical data is available, it is no longer acceptable because software projects are different in terms of development aspects. Therefore, it is better to understand the relationship between productivity and other UCP variables. This paper examines the impact of data locality approaches on productivity and effort prediction from multiple UCP variables. The environmental factors are used as partitioning factors to produce local homogeneous data either based on their influential levels or using clustering algorithms. Different machine learning methods, including solo and ensemble methods, are used to construct productivity and effort prediction models based on the local data. The results demonstrate that the prediction models that are created based on local data surpass models that use entire data. Also, the results show that conforming the hypothetical assumption between productivity and environmental factors is not necessarily a requirement for success of locality.

## 1. INTRODUCTION

Software effort estimation is an ever-present challenge in knowledge intensive industries. It is a critical prerequisite for successful scope planning, resource allocation, and scheduling during planning meetings (Ali & Gravino, 2019; Azzeh & Nassif, 2015; Gautam & Singh, 2018; Silhavy P., Silhavy P., 2019). Use Case Points (UCP) has been around for over two decades as efficient size measure, and its contribution towards accurate early-effort estimation is increasingly observed (Ali Bou Nassif, Ho, & Capretz, 2013). UCP is computed from four main variables: (1) Unadjusted Actor Weights (UAW), (2) Unadjusted Use Case Weight (UUCW), (3) Technical Complexity Factor (TCF) and (4) Environment Factor (EF). TCF variable is computed from a set of thirteen technical factors that have great influence on team performance, while the EF variable is computed from a set of eight factors that have great influence on project productivity. Each factor in both sets is assessed with a value between zero and five, in addition to a weight. The value zero means no influence, whereas value of five means strong influence, and the value of three means neither irrelevant nor strong influence. Finally, the adjusted use case points UCP is calculated by multiplying the summation of UAW and UUCW by TCF and EF as shown in Equation 1. The main challenge is how to translate the measured UCP into the most likely software effort estimate. The original UCP model that was proposed by Karner (AB & 1993, 1993) employed productivity as a second driver to generate effort as shown in Equation 2. This approach is frequently used among researchers when estimating effort from functional size measures such as Function Points or UCP (AB & 1993, 1993; Azzeh & Nassif, 2016; Mendes, 2004; A.B. Nassif, Capretz, & Ho, 2014; Ali Bou Nassif et al., 2013). Since there were no enough data to tune productivity, Karner (AB & 1993, 1993) demonstrated that the optimal productivity ratio that can be used to generate effort from UCP is 20 person-hours per one UCP.

The term productivity in software development is defined as the ratio between the amount of software produced (i.e. size) to the labor and cost of producing it (i.e. effort) (Mendes, 2004; Petersen, 2011; Rodriguez et al., 2012). This ratio is interpreted as $\frac{EFFORT}{UCP}$ in UCP effort estimation method. Thus, the smaller the productivity ratio is, the more productive the project is likely to be. Hence, the effort needed to build the project is also reduced as confirmed by Equation 2. Since the productivity is yet hard to be measured at early stage of software development, the concept Productivity Delivery Rate (PDR) is commonly used among researchers, which measures the post productivity from previous historical projects. Therefore, the terms productivity and PDR will be used interchangeably throughout this paper. However, there were substantial discussions regarding how to predict productivity when historical data is available. Some authors proposed to use multiple size measures (Mendes, 2004; Rodriguez et al., 2012). Other researchers proposed a framework to use environmental factors as indicators for measuring productivity (Azzeh & Nassif, 2017, 2016).

$$UCP = (UAW + UUCW) \times TCF \times EF \quad (1)$$

$$EFFORT = PDR \times UCP \quad (2)$$

This paper presents an insight on the performance of productivity prediction from multiple UCP size variables (i.e. UAW, UUCW, TCF and EF), using locality approaches based on environmental factors. The locality approaches refer to methods that can produce estimation from the most similar training examples to the project under estimation (Minku, Technology, & 2013, 2013). Since effort estimation datasets tend to be rather small and heterogeneous, the locality approaches are likely to be more adequate and produce better accuracy than models that do not use locality (Ekrem Kocaguneli, Kultur, & Bener, 2009). In literature, there are many approaches that can be used as pre-stage of building prediction models to produce local data such as clustering, feature selection and classification algorithms. Minko et al. (Minku et al., 2013) demonstrated that different locality approaches boost the performance of ensemble effort estimation methods. We favored using environmental factor to work as partitioning factor to obtain local datasets because there is direct impact from these factors on productivity as it has been proved in previous studies (Ali Bou Nassif et al., 2013)(Azzeh & Nassif, 2017).

The efficiency of productivity prediction based on locality of environmental factors is supposed to be influenced by the hypothetical assumption of these factors. The hypothetical assumption states that the PDR and effort should be proportionally influenced by levels of environmental factors (considering that each factor has six levels based on the scale proposed by Karner (AB & 1993, 1993)). Table1 shows the description of all environmental factors. The first six variables (E1 to E6) are supposed to have inverse proportional with PDR, which means that when any one of them is increased, the PDR is supposed to improve (i.e. the ratio of PDR becomes smaller), and the effort needed to accomplish the project is supposed to reduce. In contrast, the remaining two factors E7 and E8 are supposed to be directly proportional with PDR, which means when part time staff are hired, and programming language is difficult, the PDR and effort are supposed to increase. These assumptions made here are supposed to be true when we encounter real industrial data. This hypothetical assumption is yet not confirmed when we encounter real dataset. For example, it is acknowledged that when developers have enough excellent skills and experience in object oriented, the PDR would reasonably be reduced. Some projects in real dataset do not respect this hypothesis, so we may find data for some projects which have been developed by team members with very little experience and capability, but the measured PDR ratio is quite small. We understand that the environmental factors are not the only factors that affect measuring UCP and effort, but even that they are supposed to conform the proportional hypothesis with PDR. The impact of locality approaches on predicting productivity from multiple UCP variables has never been examined before. Therefore, this paper attempts to answer questions related to above discussion. We propose three main research questions as follows:

**RQ1.** Does the proportional assumption between environmental factors and productivity remain valid when we encounter real data?
To answer this question, we proposed the following null hypothesis:

$H_0$: Productivity is proportionally influenced with levels of each environmental factor

The analysis made in this paper shows that not all factors conform the hypothetical relationship between environmental factors and productivity which rejects our null hypothesis. Only five factors demonstrate positive relationship with PDR.

**RQ2.** Does the productivity prediction model based on UCP size metrics improved when using local data based on environmental factors?

The answer to this question can give us insight on how much the estimators are aware of the exact interpretation of these factors, and how they should be assessed. It also gives us an initial impression about performance of locality approaches based environmental factors for predicting PDR from UCP size metrics. To accomplish that we used three common machine learning and statistical regression methods to build productivity prediction models based on local data namely, Support vector regression (Valdimir, V., 1995), Stepwise regression and Regression tree (CART)(Breiman, Friedman, Olshen, & Stone, 1984). From the empirical results we found that the prediction models that are created based on local data surpass models that use entire data. Also, the findings show that conforming the hypothetical assumption between productivity and environmental factors is not necessarily a requirement for success of locality.

**RQ3.** Does the proposed ensemble learning model perform better than solo methods for productivity prediction?

In addition to the use of three solo prediction methods, we investigated the efficiency of ensemble learning based on locality for treating such problem by proposing a new ensemble approach to aggregate productivity predictions. This question is designed to investigate the stability of ensemble learning against base prediction methods and to examine the improvements that can be achieved when base methods are not able to produce accurate prediction results. Ensemble learning method aims to build a combination of prediction models that is supposed to work accurately better than base models. The ensemble learning can work well when each based model behaves differently and comes with its own assumption and configuration parameters. The final prediction in ensemble can be aggregated from a set of initial solutions by applying either simple statistical methods such as mean, weighted mean or by a more complex machine learning based methods such as Bagging and boosting. Thus, the models in ensemble can boost each other in which estimate errors can be reducing because each method in the ensemble tries to minimize and patch errors made by other methods. Based on empirical results we found that our proposed ensemble model surpasses other base models and showed that it can perform good with extracted local data.

In summary, the contributions of this paper are as follows:
1. The hypothetical assumption of productivity and environmental factors is investigated.
2. The performance of some regression models based on local data for productivity prediction are evaluated.
3. A new ensemble approach is proposed to improve the performance of base regression models.

The present paper is structured as follows: Section 2 presents related work. Section 3 describes the dataset and its statistical properties. Section 4 introduces the research methodology. Section 5 presents the results. Section 6 presents main threats to validity, and the paper ends with conclusion in Section 7.

TABLE 1 Description of Environmental factors

| Acronym | Factor Name | Description |
| --- | --- | --- |
| E1 | Familiar with RUP | it measures the experience of developer with development process |
| E2 | Application Experience | it measures the developers experience with kind of the project |
| E3 | Object Oriented Experience | it measures the developers experience in Object Oriented development |
| E4 | Lead Analyst Capability | it measures the capability, skills and knowledge of the analyst in collecting and analysis requirements |
| E5 | Motivation | it measures degree of motivation among developers |
| E6 | Stable Requirements | it measures the requirements stability |
| E7 | Part-Time Staff | it measures the level of outside developers and consultants needed to accomplish a software project. |
| E8 | Difficult Programming Language | it measures the difficulty of the programming language used |

## 2. Related Work
### 2.1 Use Case Points

Studies on UCP and effort estimation are increasingly observed. These studies can be classified into three groups: enhancement (M Ochodek, Alchimowicz, …, & 2011, 2011; Robiolo, Badano, & Orosco, 2009), evaluation (L. M. Alves, Sousa, Ribeiro, & Machado, 2013; Azzeh & Nassif, 2017; Azzeh, Nassif, & Banitaan, 2018; M Ochodek, Nawrocki, & Kwarciak, 2011; Radek Silhavy, Silhavy, & Prokopova, 2018) and prediction (Azzeh & Nassif, 2016; Satapathy & Rath, 2014; Radek Silhavy, Silhavy, & Prokopova, 2017). Majority of previous studies focused on predicting effort from UCP where different machine learning and data mining techniques were used. Kamal et al. (Kamal, Ahmed, & El-Attar, 2011) introduced fuzzy logic system to the estimation process in order to handle the imprecise information available during the early stages of software development. The authors also proposed a framework to classify and compare the approaches based on use case point and provide such aid to practitioners.

Nassif et al(Henderson-Sellers et al., 2002; Ali Bou Nassif, Capretz, & Ho, 2010, 2011; Ali Bou Nassif, Fernando Capretz, & Ho, 2011) used Mamdani and Sugeno fuzzy models to calibrate the abrupt change in the complexity weights of the use cases. On the other hand, some studies focused on evaluating the structure of UCP (Azzeh & Nassif, 2017; M Ochodek, Nawrocki, et al., 2011). They found that UCP can be simplified or enhanced to increase or maintain accuracy and avoiding insignificant variables. In this regard, Robiolo (Robiolo et al., 2009; Robiolo & Orosco, 2008) defined two metrics called transactions and paths, which capture two key aspects of software applications: software size and complexity. Transactions can be identified from the textual description of use cases. Whereas Path calculation is based on McCabe's cyclometric complexity. Alves et al. (R. Alves, Valente, & Nunes, 2013) backed up the claim that reflecting HCI concerns (the human centric models also called iUCP) provides better estimate of effort than UCP models by conducting a statistical analysis of cost estimation results between iUCP and UCP. Frohnhoff and Engels (Frohnhoff & Engels, 2008) applied UCP model on 15 industrial projects and found that the results deviated considerably. They suggested some improvements over UCP method by standardizing the weighting of use cases and providing a user guide which defines the level of abstraction and complexity of a use case. Ochodek et al. (Mirosław Ochodek & Nawrocki, 2008) (M Ochodek, Alchimowicz, et al., 2011) (M Ochodek, Nawrocki, et al., 2011) proposed a new method for measuring use-case complexity that is called TTPoints. It includes knowledge regarding semantics of transactions, numbers of business objects and interacting actors. The authors also simplified UCP model by eliminating the weight of the actors because it is insignificant with respect to the use case weights.

## 2.2 Productivity Prediction

Very few studies shed some light on the importance of productivity for predicting effort from (L. M. Alves et al., 2013; Azzeh & Nassif, 2016; Azzeh et al., 2018; Colomo-Palacios, Casado-Lumbreras, Soto-Acosta, García-Peñalvo, & Tovar, 2014; Lagerström, von Würtemberg, Holm, & Luczak, 2012; Ali Bou Nassif et al., 2013). Previous UCP effort estimation models used fixed or limited productivity values, which have been figured out from a small number of observations (AB & 1993, 1993). In the original UCP method, Karner (AB & 1993, 1993) suggested using the default productivity value which is equal to 20 person-hours/UCP. Karner found that all examined projects tend to use the same productivity value. This assumption became the basis for late models especially when there are no historical available datasets. Several studies came later criticized this assumption and demonstrated that using the same productivity for all projects irrespective of their type, complexity, and team size is not meaningful and produce bad accuracy(Azzeh & Nassif, 2017; Azzeh et al., 2018). In this direction, Schneider and Winters (SW) recommended using three levels of productivity based on examining the environmental factors (Schneider, 2001). They use Equation 2 but with three levels of productivity based on analyzing environmental factors. The principal idea is to count the number of factors from the set E1 to E6 that have influence value less than three and count the number of factors from the set E7 to E8 that have influence value larger than 3. Based on total count, the productivity is computed. If the total count is less than or equal to 2, then the efficiency is fair and assigned 20 hours/UCP. If the total count is between 3 and 4 (inclusive) then the efficiency is low and assigned 28 hours/UCP, and finally if the total count is greater than 4 then the efficiency is very low and assigned 36 hours/UCP. Azzeh et al. (Azzeh & Nassif, 2015, 2017) examined and analyzed the relationship between Environmental factors and project productivity and found that these factors can significantly help in predicting productivity at early phases. Nassif et al. (2013) did not use Equation 2, but they proposed a non-linear relationship between effort and UCP with productivity. They proposed four productivity levels based on analyzing environmental factors. The possible productivity values that can be selected are 0.4, 0.7, 1.0, 1.3 UCP/hour.

Kitchenham and Mendes (Kitchenham & Mendes, 2004) demonstrated that the productivity can be effectively predicted from multiple size measures. Rodriguez et al. (Rodriguez et al., 2012) analyzed the team size and productivity variables in International Software Benchmarking Standards Group (ISBSG) dataset in attempt to investigate their usefulness for effort estimation. They evaluate these variables when a dataset is partitioned based on type of development and programming languages. The results suggested that a team with size 9 or more is not productive, and enhancements projects have better productivity than new projects. Azzeh et al. (Azzeh & Nassif, 2017, 2016; Azzeh et al., 2018) studied and analyzed the relationship between environmental factors and productivity using some well-known estimation models. They found that environmental factors are good contributors in predicting productivity if we reduce uncertainty in evaluating environmental factors. They concluded that the productivity computation should be flexible and adjustable when data is available. The flexibility requires the productivity to be affected by UCP factors assessment. The adjustability means the productivity of one variable should be adjusted based on the productivities of the historical projects.

### 2.3 Locality Approaches for software effort estimation

Literature on examining locality approaches for software effort estimation is few. The locality approaches refer to methods that can produce estimation from the most similar training examples to the project under estimation. Since effort estimation datasets tend to be rather small and heterogeneous, these locality approaches are likely to be more adequate and produce better accuracy than models that do not use locality (Gallego, Rodríguez, Sicilia, Rubio, & Crespo, 2007; Ekrem Kocaguneli et al., 2009; Minku et al., 2013; R Silhavy, Silhavy, Technology, & 2018, 2018). There are various kinds of locality approaches such as clustering, feature selection and classification algorithms. Multiple studies reported improvements when considering locality with their prediction models. For example, Gallego et al. (Gallego et al., 2007) used Expectation Maximization clustering method to separate training examples into more homogeneous examples, then they construct regression models for each cluster. Kocaguneli et al. (Ekrem Kocaguneli et al., 2009) constructed multiple decision tree models based on different training examples. Minko et al. (Minku et al., 2013) studied the impact of locality on the construction of ensemble methods. They found that locality approaches such as clustering algorithm can help to improve accuracy of ensemble effort prediction models.

Silhavy et al. (Radek Silhavy et al., 2018) examined various data points selection algorithms for UCP effort estimation using UCP size metrics. They employed various clustering algorithm such as k-means, Gaussian mixture and moving windows to build accurate effort estimation model using Multiple Linear Regression and based on UCP size metric. The main differences between our paper and Silhavy's work are: Our work focusses mainly on productivity prediction then effort estimation from UCP and productivity, while Silhvay work focus on direct effort estimation from four UCP metrics. The locality approach that is constructed in our work is completely different than the data points selection used by Silhavy. In our work we used environmental factors that are supposed to have relationship with productivity to divide whole dataset to coherent subset of data sets that are most likely boosting productivity prediction from UCP metrics and consequently effort estimation based on UCP. Furthermore, we used various machine learning algorithms such as support vector machine and regression tree in addition to ensemble learning to construct various productivity prediction models. In Silhavy's work they used existing data mining algorithms such us k-means and gaussian mixture algorithms to divide entire data set into local datasets and construct local linear regression between effort and UCP size metrics. However, there is no empirical study that examines the impact of locality approaches for predicting productivity and effort from UCP variables.

### 3. DATASET DESCRIPTION

The dataset used in this paper was collected by Nassif et al. (2013). The dataset includes both industrial and educational projects. There are 45 industrial projects developed by a company for information systems projects such as multi-warehouses bookstores, chains of hotels, multi-branch universities and pharmacies. The architectures used to develop these projects are 2-tier desktop application and 3-tier web architecture. Object Oriented Design was used and Sybase Power Designer 12.5 and 15 was utilized as a CASE tool. Use case diagrams with use case descriptions were designed to be part of software requirements specification (SRS). Because of confidentiality reasons, we could not have access to the SRS of all projects. However, a questionnaire was prepared that had detailed information about the actors and use cases in the UML diagram such as the number of transactions and complexity of each use case. The questionnaire was completed by trained software developers in the company and these details were collected in order to calculate the software size in use case points (UCP) metrics. Likewise, there are 65 educational projects were collected from 4th year software engineering projects and Master's student projects and theses from a university in Canada. The projects were developed and implemented using UML and object-oriented languages. Use case diagrams with use case descriptions were also available in the SRS and this helped calculating UCP software size. Some students were working as part time and some as full time on their projects/theses. This variation was taken into consideration as explained in Table 2. Table 2 shows the descriptive statistics of 110 projects.

At the first glance, we can observe that most variables do not follow normal distribution because their Skewness and Kurtosis values are not zero. Negative Skewness means that the data is skewed to left and positive Skewness means that the data is skewed to right, while zero means that the data is normally distributed. Kurtosis measure describes the tail shape of data distribution. Negative Kurtosis represents thin-tail, positive Kurtosis represents fat-tail, while zero represents standard tail shape. The effort variable has large standard deviation which might be happened because of the presence of outliers.

TABLE 2 Descriptive statistics of UCP dataset

| Variable | Mean | StDev | Skewness | Kurtosis |
|---|---|---|---|---|
| PDR | 18.07 | 4.5 | 0.2 | -0.22 |
| Effort | 9405.00 | 31487.0 | 5.2 | 28.48 |
| UCP | 369.8 | 1044.9 | 5.06 | 26.5 |
| UAW | 19.25 | 5.5 | -0.25 | -1.13 |
| UUCW | 375.00 | 1123.0 | 4.93 | 25.4 |
| TCF | 0.97 | 0.064 | 0.31 | 0.51 |
| EF | 0.96 | 0.14 | 0.07 | -0.32 |

Figure 1 shows the histogram of PDR variable. We can notice that the PDR variable is slightly skewed to right as its skewness is equal to 0.2 and its distribution is flatter than normal with a kurtosis value of -0.22. Generally, the projects that are skewed to the right are more productive than those are skewed to left. This is an indication that these projects possess better productivity and consequentially need less effort. Although this is quite intuitive from the histograms, it was also confirmed by with Kolmogorov–Smirnov test for normality. An important observation that can be drawn up from Figure 1, is that the variability of the PDR spread. This variability rejects the assumptions made by Karner (AB & 1993, 1993; Robiolo et al., 2009) and SW regarding the static value of productivity.

Next, we study the relationship between PDR variable and other UCP variables as shown in Table 3. The analysis was made based on using non-parametric analysis Spearman's rank correlation method. First, we can observe that there is good association between PDR and EF, UUCW variables. This confirms the assumption made in some studies that the environmental factors are good indication for the amount of productivity needed to develop software project when no historical datasets are available (Azzeh & Nassif, 2017). On the other hand, PDR has negative weak association with UAW, and no correlation with TCF.

TABLE 3 Spearman's Rank Correlation between PDR, and UCP variables

| Variable | r | p-value |
|---|---|---|
| UAW | -0.148 | 0.122 |
| UUCW | 0.456 | 0.001 |
| TCF | 0.006 | 0.95 |
| EF | 0.386 | 0.000 |

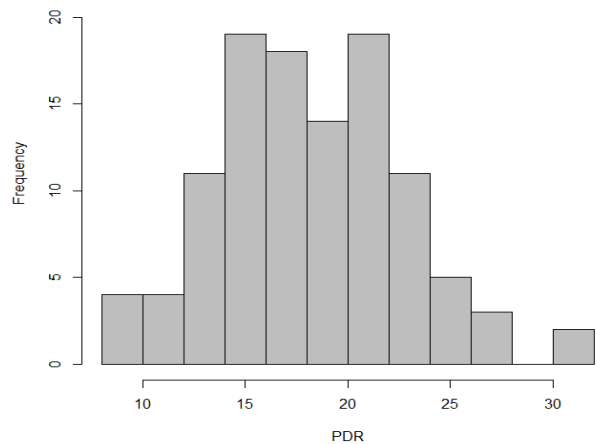

**Fig 1.** Histogram for PDR

## 4. Methodology

The first part of our analysis aims at empirically validating the hypothetical relationship between influential levels of each environmental factor and PDR. To examine the hypothetical relationships (RQ1), the interval plot with 95% significant analysis was used, which shows the confidence interval for the mean of the PDR.

The interval plot with 95% confidence is used for two reasons:
1) It is a graphical data interpretation technique which allows us to visually assess the difference between multiple groups and build provisional decision.
2) Since we are interested to investigate the significant differences between means of different populations (i.e. different levels) the point mean estimate alone is not enough to judge that multiple subsequent of distribution population are significantly different. It means that for each such sample, the mean, standard deviation, and sample size are used to construct a confidence interval representing a specified degree of confidence, say 95%. Under these conditions, it is expected that 95% of these sample-specific confidence intervals will include the population mean (Masson & Loftus, 2003)

The second part of our analysis aims at applying locality approaches based on environmental factors to build multiple productivity prediction models from UCP size variables as shown in Figure 2. This part facilitates answering research questions RQ2 and RQ3. The next subsections explain the necessary steps that are required to build productivity prediction models based on local data points. The complete replication package of our experiments can be found at: https://cutt.ly/xjwVodC

### 4.1 Data preprocessing

Data preprocessing is an important step in machine learning applications. The first step is to check and eliminate observation that have outliers in one or more features. Most of software effort estimation datasets frequently having outliers, and UCP has no exception. We used Z-score measure to detect outliers. Z-score is a parametric outlier detection method that indicates how many standard deviations a data point is from the sample's mean. Two observations were eliminated because they have outliers in some features. The next important issue is to ensure that there are no missing values. Fortunately, there were no missing values in the dataset. All variables are then normalized using min-max approach to have the same influence on the output variable.

### 4.2 Evaluation measures

The accuracy of the prediction models is evaluated based on three common accuracy measures. These measures are: Mean Absolute Errors (MAE), Mean Balanced Relative Error (MBRE), and Mean Inverse Balanced Relative Error (MIBRE) as shown in Equations 3, 4 and 5. They have been selected from literature because they are frequently used and they have less bias than other accuracy measures such as Mean Magnitude Relative Error (MMRE) (Foss, Stensrud, Kitchenham, & Myrtveit, 2003; Shepperd, Technology, & 2012, 2012).

$$MAE = \frac{\sum_i^n |e_i - \hat{e}_i|}{n} \qquad (3)$$

$$MBRE = \frac{1}{n} \sum_1^n \frac{|e_i - \hat{e}_i|}{min\,(e_i, \hat{e}_i)} \qquad (4)$$

$$MIBRE = \frac{1}{n} \sum_1^n \frac{|e_i - \hat{e}_i|}{max\,(e_i, \hat{e}_i)} \qquad (5)$$

Where $e_i$ and $\hat{e}_i$ are the actual and estimated effort of $i^{th}$ project.

### 4.3 Locality approaches

As it was shown in section 3, the whole dataset is described by seven variables: UAW, UUCW, EF, TCF, UCP, PDR and Effort. Therefore, the local datasets are also described by the same variables. There are various kinds of locality approaches such as clustering algorithms, feature selection and classification algorithms (Minku et al., 2013). In our case, the situation is different, we do not use locality approaches directly, but we use influential levels of environment factors (E1 to E8) to partition training examples into more homogeneous examples that are similar with respect to influential levels in each factor. Therefore, two types of locality were used, the first one is to partition whole data based on each influential level of each environment factor. The second approach is to use k-means clustering to partition whole data based on environment factors.

#### 4.3.1 Locality Approach 1: Levels of environmental factor

In this approach, the whole data is partitioned based on influential levels of each environment factor. In other words, if factor E1 is considered as partitioning factor then the whole dataset is divided to $n$ local datasets based on number of levels in that factor. The value $n$ represents number of discrete values in the environmental factor which usually range from 1 to 5. But, since some levels do not contain enough number of observations, we merge two adjacent levels together for sake of predictions as it is discussed in section 5.

#### 4.3.2 Locality Approach 2: k-means

The second approach aims at partitioning the whole data based on the concept of clustering all environment factors. To achieve that, we use k-means to partition the whole dataset into local datasets based on environmental factors only. K-means algorithm belongs to the family of similarity-based clustering techniques that use the default distance measure (i.e. Euclidean distance) to identify closest objects. The k-means algorithm requires initially determining the optimal number of clusters that can separate observations efficiently. Therefore, we use Dunn Index as measure of separation among clusters. The number of clusters is chosen empirically by changing $k$ from 2 to 10, then the $k$ value that produces larger inter distance among cluster centers and smaller intra distance among observations in the same cluster is chosen.

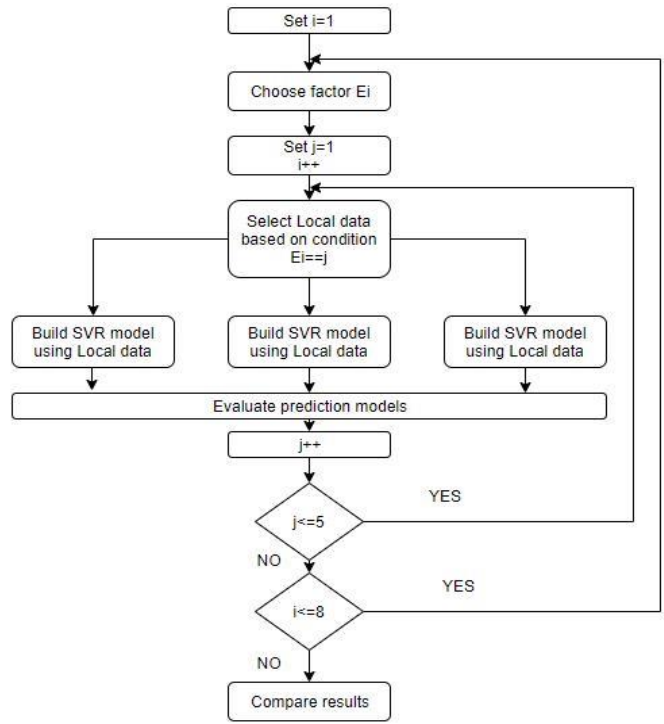

Fig. 2 Flow Chart of our research methodology. The $i$ and $j$ variables represent indexes of environmental factor and influential level respectively.

### 4.4 Choice of Learning Machines

For each local dataset, the productivity prediction models are constructed using the following most common regression methods: Regression Tree (CART)(Breiman et al., 1984), Support Vector Regression (SVR)(Valdimir, V., 1995), Stepwise Regression (SR). These methods were chosen because they are related to locality, and they will be called later base model.

CART is a machine learning algorithm that can be used to both regression and classification. CART can be viewed as object separation rules based on data feature space where each branch separates training data with similar characteristics therefore the prediction of test object is produced from local data the belongs to that branch. So, CART itself is based on notion of locality (Minku et al., 2013). The main advantage of CART over clustering is that it is created by considering not only the existing input features of the training objects, but also the impact of their values on the output.

SR is a multiple linear regression model that is created gradually using forward or backward feature selection of training objects. For backward selection, the regression model is stared with all input features then in each step the least significant feature is removed until none meet the criterion. On the other hand, the forward selection starts with the most significant feature and iteratively add the most significant feature from those that are not in the model. In both cases the final model ends with the most significant features. In this paper the backward selection has been favored over. it is important to make sure that assumptions related to using SR are not violated such as data skewness and distribution. In each validation step the normal distribution of each feature have been checked. If any feature is not normally distributed, then the feature values are transformed to a natural logarithmic scale to approximate a normal distribution (Mendes, Watson, Triggs, Mosley, & Counsell, 2003).

The SVR is constructed based on the original support vector machine that was mainly proposed for classification. The SVR creates a hyperplane that separates the data points into two classes with maximal margin. The margin is defined as the distance between hyperplane and nearest points. The accuracy of SVR depends on the choice of kernel functions that are used to construct optimal hyperplane. There are various linear and nonlinear kernel can be used with SVR such as polynomial, sigmoid and radial basis function. The Epsilon-SVR has been used with radial basis function as kernel function to achieve locality as radial basis function is implemented based on the idea of similarity. In addition, the sequential minimal optimization method is used to find the best hyperplane. The memory consumption is controlled by the value that specifies the kernel matrix cache with 5000. The stopping criteria is set by 0.001. In this paper we used radial basis function because it helps.

All chosen learning methods have advantage that they only use the significant independent variables in their model construction, which is considered a type of another locality approach called feature selection. All prediction models were implemented in RStudio using R language. Equation 6 shows the mathematical model of productivity prediction model that is implemented based on local data, using four basic UCP metrics. However, the productivity of new project is predicted from the class that is most similar to it. The effort of new project is estimated using predicted productivity from previous step and the measured UCP as shown in Equation 2.

$$Productivity = f(UAW, UUCW, TCF, EF) \qquad (6)$$

Where *f* is the constructed base productivity prediction model by one of the employed machine learning algorithms.

### 4.5 The proposed ensemble method

In addition to the basic three regression models, we proposed in the previous section, a new ensemble approach based on three base models mentioned above. The basic idea of the ensemble model is to construct each base model on the same training examples, then the final productivity prediction is obtained weighted average of the predictions obtained from each base model. The objective of this step is to compute weight of each local base model based on its *MAE*, *MBRE* and *MIBRE* of training data. First, the training data is used to construct various local base models using bagging algorithm. In other words, the training data is divided into many subsets of training and testing data using leave one cross validation to build local models. Then the *MAE*, *MBRE*, *MIBRE* of is computed. Then the evaluation errors of all models are normalized individually using min-max approach to have the same influence. The symbols $\overline{MAE}, \overline{MBRE}$, and $\overline{MIBRE}$ represent the normalized error measures. These normalized values are then entered into discounting function to compute weight of each base model as shown in equations 7, 8 and 9. This discounting function is implemented using sigmoid function as shown in Figure 3. The model with small errors will get high weight while models with high error rate will get less weight. The α parameter in the equations represents a scaling factor to make weight is close to 1 when any normalized error≃0. However, we use α=15 after several experiments. Later, the average of weights of three evaluation measure are calculated to compute the corresponding weight of the model as shown in Equation 10. The final productivity prediction is computed as shown in equation 11 by applying weighted average.

$$w^i_{MAE} = \frac{1}{1 + e^{\alpha(MAE_i - \overline{MAE})}} \tag{7}$$

$$w^i_{MBRE} = \frac{1}{1 + e^{\alpha(MBRE_i - \overline{MBRE})}} \tag{8}$$

$$w^i_{MIBRE} = \frac{1}{1 + e^{\alpha(MIBRE_i - \overline{MIBRE})}} \tag{9}$$

$$w_i = \frac{w^i_{MAE} + w^i_{MBRE} + w^i_{MIBRE}}{3} \tag{10}$$

Where $w^i_{MAE}$, $w^i_{MBRE}$ and $w^i_{MIBRE}$ are weight obtained from normalized error measure *MAE*, *MBRE* and *MIBRE* for model *i*.

$$productivity_j = \frac{\sum_{i=1}^{3} productivity_i \times w_i}{\sum_{i=1}^{3} w_i} \tag{11}$$

Where $productivity_j$ is the aggregated productivity of the test project. $productivity_i$ is the productivity of test project obtained by model *i*.

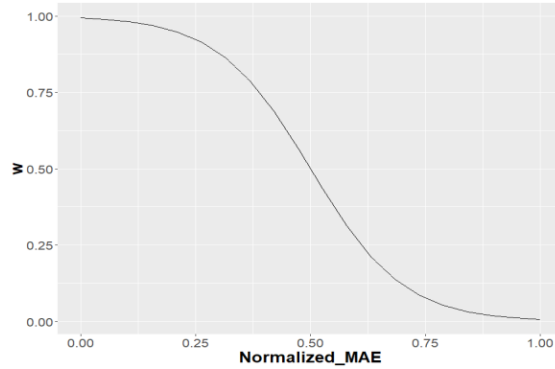

Figure 3. Sigmoid discounting function with $\alpha = 15$

### 4.6 Validation approach

All models are trained and tested using leave-one out cross validation. In each run, a single object is chosen as test project while the remaining projects work as training dataset. Note, the prediction model is constructed based on training project. This process is repeated until all objects act as test data points. During this process the error values are calculated and then aggregated at end of the evaluation procedure. This validation procedure is robust and ensure that all objects are validated equally without bias (E Kocaguneli, Software, & 2013, 2013b).

# 5. RESULTS

Using homogeneous or heterogeneous dataset has a great influence on the construction of productivity and effort prediction models. Effort estimation models are easily influenced with the structure of data. Therefore, it is important to investigate the stability of predictions in term of both homogeneous and heterogeneous data. This section presents the empirical results on the performance of locality approaches for productivity and effort prediction. The following subsections present answers to our proposed questions through statistical and empirical validations as described in Section 4.

## 5.1 RQ1. Investigating proportional assumption of PDR and environmental factors

This section is designed to answer RQ1 which is about investigating the hypothetical proportional assumption between PDR and environmental factors. It is widely acknowledged that the environmental factors effectively contribute in improving productivity predictions. But, in order to ensure that each factor has positive influences on the productivity, the hypothetical relationship between PDR and influential levels of environmental factor must be respected. If all factors do respect the hypothetical relationship with PDR, then we can comfortably ensure that these factors can partition the entire dataset into local homogenous datasets. To clearly answer this question, we first study the distribution of PDR variable across each influential level of every environmental factor. In other words, for each environmental factor, the PDR values are divided into five groups corresponding to the number of levels. Then the 95% confidence interval plot of PDR values in each level is drawn as shown in Figures 4 to 11. The interval plot shows the confidence interval for the mean of the PDR. The first six variables (E1 to E6) are supposed to have inverse proportional with PDR, which means that when developers experience and team capabilities are increased the PDR is reduced (i.e. improved), and the effort needed to accomplish projects is supposed to reduce. In contrast, the remaining two factors (E7 and E8) are supposed to have direct proportional with PDR, which means when part time staff are hired, and programming language is difficult, the PDR and effort are supposed to increase. These assumptions made here are supposed to be always true when we encounter real data. The purpose of this analysis is to investigate whether the collected real projects conform the hypothetical relationship between productivity and environmental factors.

The typical situation when evaluating factors E1 to E6 in Figures 4 to 11 is that the mean of PDR is decreased when the level of any of these factors is increased, thus the minimum, maximum and mean of the interval plot should also be reduced. The larger the factor value is, the smaller PDR is likely to be. If we look at 95% confidence interval plots for the mean of PDR in Figure 4, we can see that the trend of projects partially conforms the hypothetical assumption. Mainly, there is a decrease trend in interval plots beginning from level 2 to level 5. Mean of PDR values in level 1 is not larger than mean of PDR in level 2 which is the odd case in E1 factor. In our case, we can observe that the interval plots of level 1 is quite large due to small number of observations in this level. The most interesting observation is that there is an overlap between every two adjacent interval plots, which confirms that these adjacent levels are not significantly different. Usually, the larger the sample size, the smaller and more precise the confidence interval is. Similarly, the trends in Figures 5, 7 and 9 for factors E2, E4 and E6 respectively follow the same tendency of Figure 4, but in Figure 5 the odd case is in level 5 which is supposed to be less than the mean of PDR in level 4. Figure 8 shows the typical distribution of PDR over levels of E5 factors. We can observe reduction in the mean of PDR as long as E5 value is increased. However, the assumption of factor E3 in Figures 6 is quite turbulent with a large overlap and no decreasing trend. On the other hand, the typical situation for factors E7 and E8 is that the 95% confidence mean of PDR is increased as long as levels of factor increased. The trend in Figure 11 for factor E8 is quite turbulent than trend of Factor E7 in Figure 10 where both factors do not confirm the hypothetical assumption. This behavior explains the misunderstanding in interpreting these two factors.

Another observation is the variability in each level of every environmental factor. Some levels have large variability than other levels in the same environmental factors. For example, for factor E1 in Figure 4, we can observe that the variability of PDR in levels 1 and 5 are much larger than variability in levels 3 and 4. The same trend is appeared in all factors. Usually, the length of confidence interval is affected by sample size so the larger the sample size, the smaller and more precise the confidence interval. This may explain the existing problem in understanding the relationship between environmental factors and PDR.

Based on this analysis we can generally conclude the relationship between PDR and influential levels in environmental factors are conformed only by one factor which is E5, and quietly conformed by four factors E1, E2, E4 and E6. Other factors did not show clear trend with respect to the relationship with the PDR variable. The main reason behind this discrepancy in the relationships refers to the fact that estimators are not aware of the correct interpretation and real impact of these factors on the development process. Also, the uncertainty associated with measurement can easily affect the accuracy of their judges. Yet, there is no empirical evidence if the factors E1, E2, E4, E5 and E6 can help in obtaining local datasets. Next subsection examines the empirical evidence of using these factors in generating local homogeneous datasets, thus better prediction models.

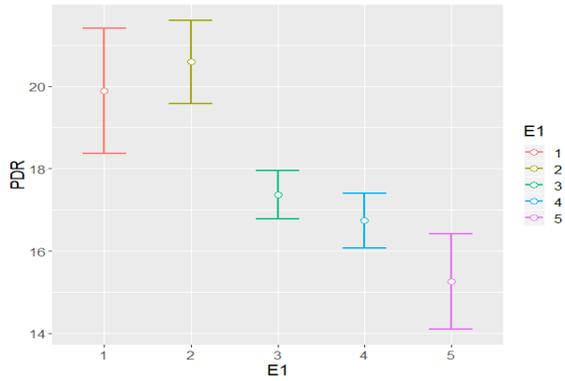
Fig. 4 Interval plot of PDR vs. E1 levels

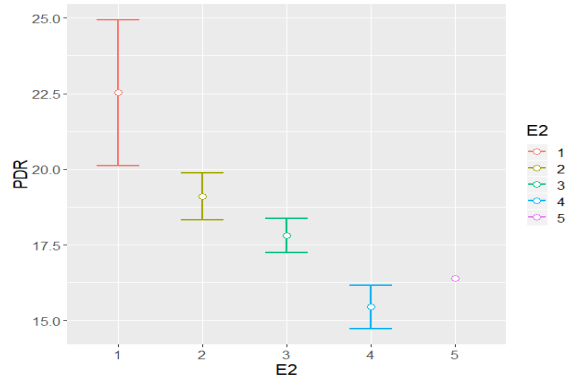
Fig. 5 Interval plot of PDR vs. E2 levels

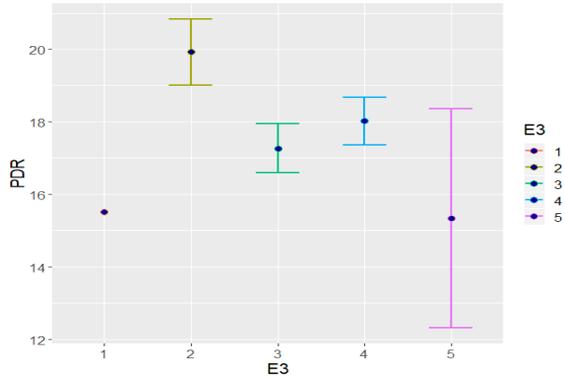
Fig. 6 Interval plot of PDR vs. E3 levels

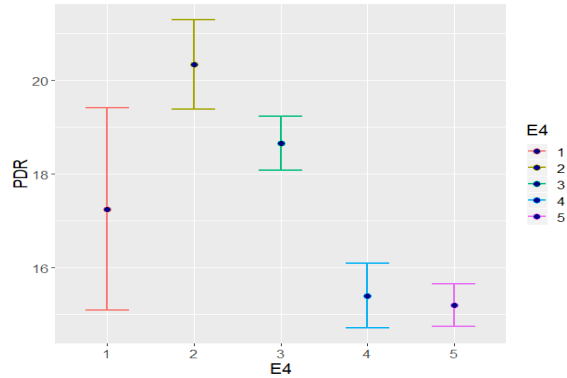
Fig. 7 Interval plot of PDR vs. E4 levels

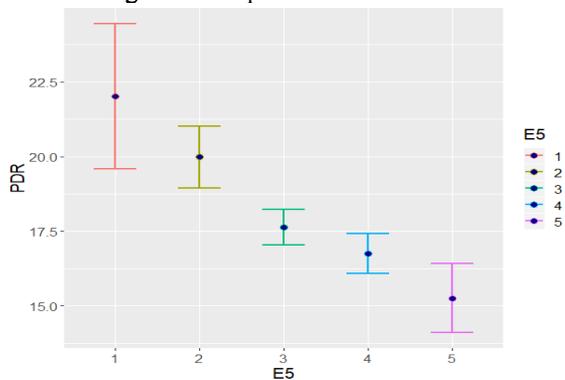
Fig. 8 Interval plot of PDR vs. E5 levels

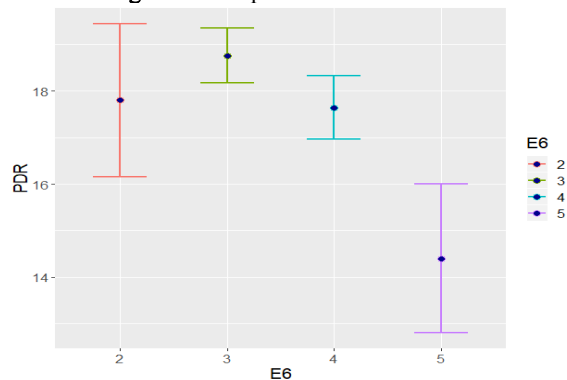
Fig. 9 Interval plot of PDR vs. E6 levels

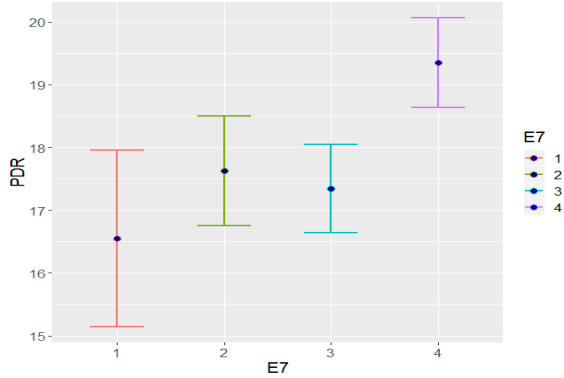 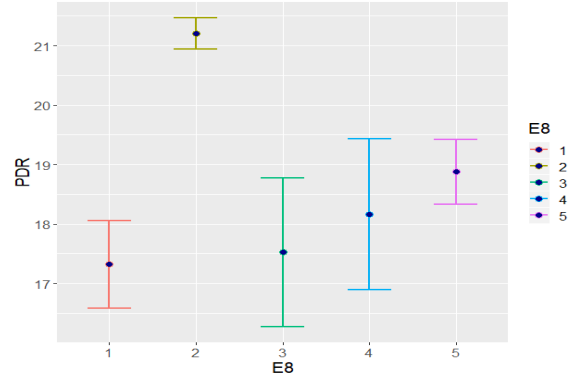

**Fig. 10** Interval plot of PDR vs. E7 levels  **Fig.** 11 Interval plot of PDR vs. E8 levels

### 5.2 Prediction accuracy of base prediction models and ensemble based on local data

This section is designed to answer RQ2 and RQ3 which are related to the prediction accuracy of each base model and the proposed ensemble model. We have seen from the previous section that one factor conforms the hypothetical relationship and four factors quietly conforms the hypothetical relationship to productivity. Therefore, we expect that making local homogeneous data based on the influential levels of these five environmental factors would contribute to better productivity prediction from multiple UCP variables than using the remaining environmental factors. Yet, we have not found any empirical evidence telling that the local data extracted based on levels of environmental factors are good enough to help in building accurate productivity and effort prediction models. To answer this question, the levels that exist in each environmental factor are used to partition a whole dataset into local datasets. Then, for each local dataset, productivity prediction models are constructed using well-known machine learning methods such as SVR, CART and SR and the proposed ensemble model. The test project is predicted from the dataset that most belong. Finally, the effort is estimated from UCP and predicted productivity using Equation 2. However, since some levels might have very few training examples that affect constructing productivity prediction models as shown in Figures 12 to 19, we decided to reduce five levels to three levels with sufficient number of observations in each level. Figures 12 to 19 show bar chart of levels in each factor. The general trend in these figures demonstrated that number of projects in the boundary levels are few. Therefore, the projects in levels 1 and 2 of each factor are combined and replaced by new level called L12. Similarly, the projects in level 4 and level 5 are combined and replaced by new level called L45. Level 3 is kept without change because this level contains sufficient number of observations in comparison with other levels. Furthermore, the k-means clustering method is used to partition the whole dataset into local datasets based on environmental factors as explained in section 4. In other words, the k-means algorithm will use all environmental factors values (E1 to E8) to partition whole dataset into local datasets. As known, the k-means algorithm requires initially determining number of clusters. The number of clusters has been decided empirically for each validation run by changing *k* from 2 to 10, then we use Dunn Index to measure the validity of clustering. The *k* value that can achieve larger inter distance between centers and smaller intra distance between observations within the same cluster is chosen. Due to characteristics of local datasets the number of clusters in each run might be changed.

The overall accuracy results are aggregated for each factor as shown in Table 4. The accurate results across each partitioning factor are in italic and red color, whereas best accuracy for each model is highlighted with light gray. We can observe that accuracy of using locality approach based on influential levels of each environmental factor is generally better than using clustering of the whole environmental factors. This leads to conclude that environmental factors contribute efficiently to partition heterogeneous dataset to local datasets. The accuracy results in general are acceptable but they are close and there is no clear trend for which method performs better than other over all environmental factors. However, if we look at the results of MAE, we can notice that our proposed ensemble model generates better accuracy over all different partition parameters. The factor E8 is the most accurate partitioning parameter for SVR, CART and Ensemble, while partitioning with factor E3 is the most accurate one for SR. The results of MAE over all models suggest that the SR based on E3 produces the smallest error value. While the results of MBRE and MIBRE suggest that ensemble model based on E8 factor are the most accurate results. With respect which factor lead to smallest MAE and MBRE values for SVR, CART and ensemble we can notice that E8 is the most accurate one. For SR the factor E3 is the most accurate one. In terms of MIBRE the results are similar to previous ones but with one different is that E4 would be the most interesting factor that can be used to partition whole dataset into local datasets for CART model. The accuracy measure MIBRE is not informative and cannot judge exactly which factor would be more accurate, but we can confirm that factor E8 would be more appropriate for most models. Surprisingly, the factors that have been found to conform the hypothetical assumption with PDR (i.e. E1, E2 and E5) did not help in obtaining local data for productivity prediction. Specifically, factors E5 (Motivation) and E7 (Part-Time Staff) do not have the ability to partition the whole dataset into more homogeneous local data. The Motivation factor is a relative measure and subject to large degree of uncertainty. On the other hand, the definition of part-time staff factor is quietly unclear, and experts exposes difficulties in choosing the correct scale for this factor.

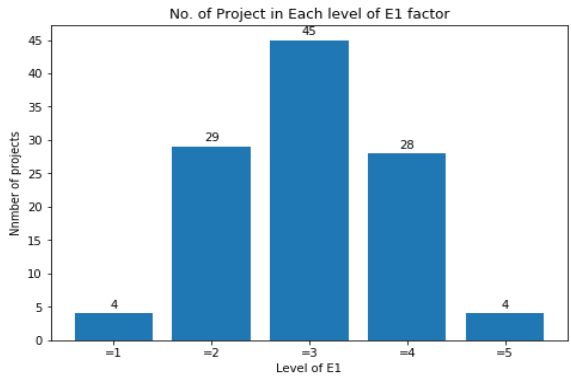

Fig. 12 Bar chart for factor E1

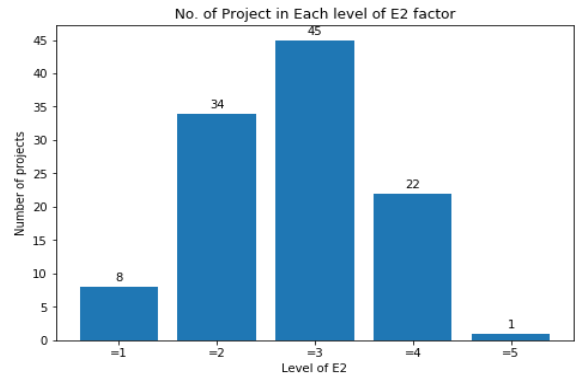

Fig. 13 Bar chart for factor E2

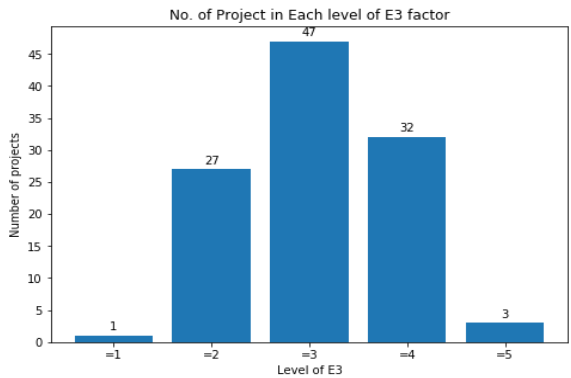

Fig. 14 Bar chart for factor E3

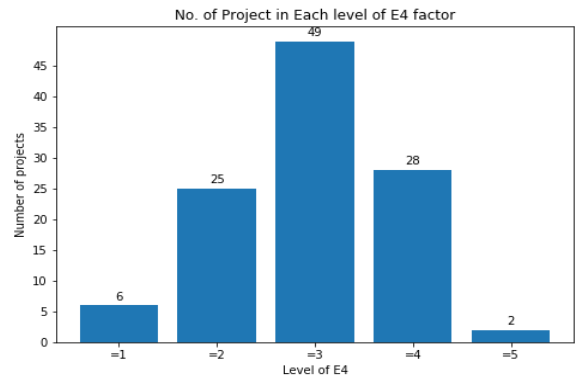

Fig. 15 Bar chart for factor E4

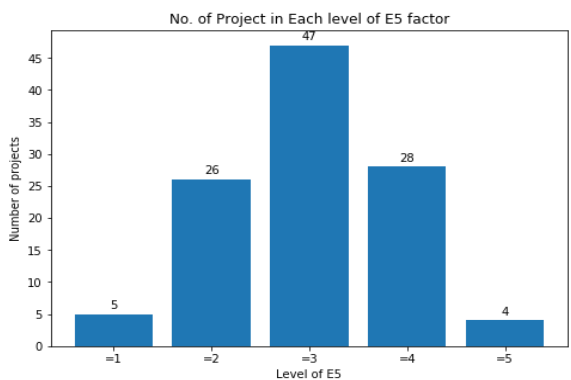

Fig. 16 Bar chart for factor E5

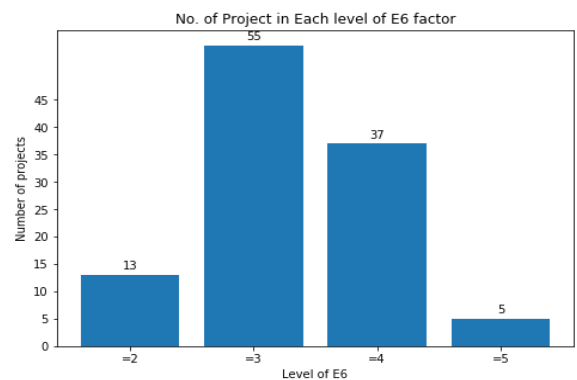

Fig. 17 Bar chart for factor E6

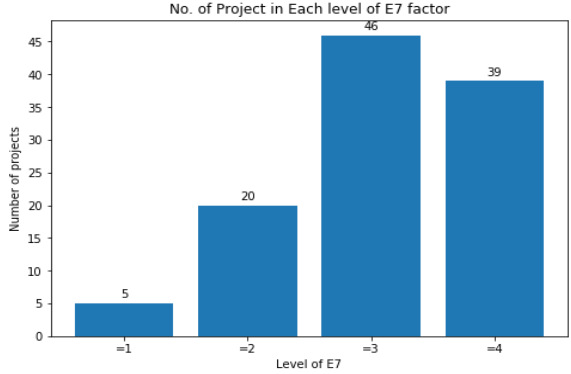
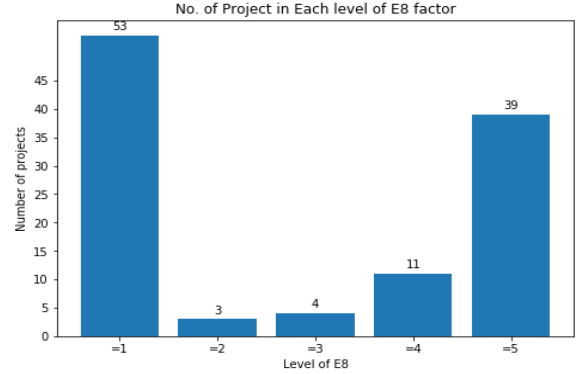

Fig. 18 Bar chart for factor E7    Fig. 19 Bar chart for factor E8

Surprisingly, if we compare between findings in Figures 4 to 11 and the recent accuracy results, we can observe that factor E5 would comfortably respect the hypothetical relationship between environmental factors and PDR, but it has not been favored by any model. While the factors E3 would help better in productivity prediction based on locality even though they did not respect the hypothetical relationship with productivity. So, we can conclude that the relationship between environmental factors and productivity is not a measure about the success of locality.

TABLE 4 accuracy results of all estimation models when applying locality approaches

| Partitioning Factor | SVR | | | SR | | | CART | | | Ensemble | | |
|---|---|---|---|---|---|---|---|---|---|---|---|---|
| | MAE | MBRE | MIBRE | MAE | MBRE | MIBRE | MAE | MBRE | MIBRE | MAE | MBRE | MIBRE |
| E1 | 2491.20 | 0.30 | 0.20 | 2080.78 | 0.29 | 0.19 | 2641.18 | 0.27 | 0.19 | *1764.37* | *0.27* | *0.17* |
| E2 | 2676.94 | 0.32 | 0.21 | 1843.52 | 0.29 | 0.19 | 2389.32 | *0.26* | 0.18 | *1713.79* | 0.27 | *0.17* |
| E3 | 2612.92 | 0.29 | 0.19 | *1071.50* | *0.25* | *0.17* | 2721.63 | 0.28 | 0.19 | 2010.36 | 0.26 | 0.18 |
| E4 | 2685.96 | 0.28 | 0.18 | 2198.79 | 0.27 | 0.18 | 2631.80 | 0.25 | 0.17 | *1733.93* | *0.22* | *0.16* |
| E5 | 2650.55 | 0.30 | 0.20 | 2494.15 | 0.29 | 0.19 | 2740.86 | 0.28 | 0.19 | *1566.25* | *0.27* | *0.18* |
| E6 | 2842.95 | 0.25 | 0.17 | 16093.45 | 0.38 | 0.20 | *3103.26* | *0.25* | 0.18 | 5725.03 | 0.26 | *0.17* |
| E7 | 2634.74 | 0.29 | 0.19 | 2435.58 | 0.30 | 0.20 | 2470.22 | 0.28 | 0.20 | *1564.37* | *0.27* | *0.19* |
| E8 | 1740.5 | 0.22 | 0.17 | 1786.7 | 0.26 | 0.19 | 1967.2 | 0.23 | 0.18 | *1486.7* | *0.21* | *0.16* |
| k-means | 3957.06 | 0.34 | 0.22 | 3515.24 | 0.31 | 0.20 | 4182.66 | 0.30 | 0.20 | *2791.24* | *0.28* | *0.18* |

Table 5 shows the accuracy results for all models without locality. The cells with boldface font represent the best value. The overall results suggest that our proposed ensemble performs better than other variants of effort estimation-based productivity. Furthermore, the productivity prediction models without locality datasets are slightly worse than the models with locality approaches. The Karner and SW models produce the worst performance in comparison to regression models. The SW model produces the worst estimates based on three datasets with large error deviation as confirmed in MAE results.

TABLE 5 accuracy results of all estimation models when not applying locality approaches

| Model | MAE | MBRE | MIBRE |
|---|---|---|---|
| Karner | 2918.05 | 0.31 | 0.23 |
| SW | 2856.69 | 0.33 | 0.25 |
| Ensemble | *1964.48* | *0.26* | *0.18* |
| Azzeh | 2664.8 | 0.30 | 0.21 |
| SVR | 2624.47 | 0.29 | 0.21 |
| SR | 2430.59 | 0.28 | 0.20 |
| CART | 2644.04 | 0.27 | 0.20 |

## 6. THREATS TO VALIDITY

The threats to validity of this study are divided into internal, construct and external threats. The main internal threat is the reliability of data measurement. Although that some of data was collected in industry, the process of translating use case diagrams into values is subject to the experience of developers and managers. Specifically, the inherent uncertainty at early phases may significantly affect measuring technical and environmental factors. Evaluating and interpreting the impact of program difficulty and part time factors is not easily understood among practitioners which may yield bias in the measured values. Therefore, we may raise a concern that the less clear the definition of the factor is, the more bias could be in its measurement. On the other hand, there is much debate about quality of data that was collected by students, which may affect the accuracy of our finding. Even though, we favor using them as collecting data from student programs is helpful in many disciplines such as education. Moreover, small programs that are developed by one developer (such as projects in student database) are of interest to many practitioners as confirmed by Personal Software Process (PSP) (Humphrey, 2002).

The main construction threat is the choice of validation approach. In this paper the leave one out cross validation is used to validate the prediction models. Although this approach has not been favored in some areas, we used this approach for two reasons: first, it has been recommended by (E Kocaguneli, Software, & 2013, 2013a) who stated that the software effort estimation should be assessed via leave one cross validation. Second, this approach has been commonly used in the majority of effort estimation studies which facilitates the comparison with these studies. Another threat is the choice of adjustment factors, we have just examined the environmental factors while other technical factors were ignored. This may reduce the ability of productivity prediction model to generate more robust results. Concerning external validity, i.e. the ability to generalize the obtained findings of our comparative studies, we used one industrial dataset and one academic dataset. Other public datasets do not offer the values of environmental factors which makes running our experiments is impossible.

## 7. CONCLUSIONS

This paper studied the impact of locality approaches on predicting productivities from multiple UCP size measures. Previous studies on effort estimation models used the same independent variables in both in classification and prediction. Unlike other locality approaches, we used environmental factors as classification parameters of local data, while the productivity prediction models are constructed from other UCP size metrics. Constructing multiple productivity prediction models from local datasets are likely to be more adequate and produce better accuracy than models that do not use locality because software datasets tend to be rather heterogeneous. This has great implication on the future of effort estimation especially when using large dataset like ISBSG.

The first part of our analysis is designed to test the hypothesis mentioned in RQ1 which states:
$H_0$: Productivity is proportionally influenced with levels of each environmental factor.
The analysis made in section 5.1 shows that not all factors conform the hypothetical relationship between environmental factors and productivity which rejects our null hypothesis. Only five factors demonstrate quite positive relationship. The main reason behind this discrepancy in the relationships refers to the fact that estimators are not aware of the correct interpretation and real impact of those factors on development team. Also, the uncertainty associated with measurement can easily affect the accuracy of their judges. Therefore, there is a need to mitigate such uncertainty and develop a complete guideline that explain how the environmental factors should be assessed and their implication on productivity and effort.

In the second part of our study, different empirical studies have been constructed based on the local data. The overall results and comparisons are promising and suggest that the models that use locality based on environmental factors are more accurate than models that do not use locality. Specifically, our proposed ensemble model surpasses other base models. Interestingly, we found that the factors that conform the hypothetical relationship between environmental factors and productivity are not able to yield local data that help in building accurate prediction models. This can explain that conforming the hypothetical assumption is not a requirement of obtaining accurate productivity prediction models from local data. Finally, using locality approaches based on environmental factors levels tend to be more accurate than using locality based on clustering of environmental factors.

UCP method has several practical implications in the fields of Project Management and Requirements Engineering. The main implication is that managers can use UCP size metrics to predict productivity and consequently the project effort. Therefore, instead of using all available data, managers can use only the most similar data that share relatively the same environmental factors values. The good news is that measuring environmental factors at early phase is not a complex process and does not need professional experience. However, to make this feasible, there should be a guide that describes in detail how the environmental factors must be assessed, which can reduce uncertainty in measuring these factors. The second implication is that using UCP method, managers can overcome the problem of underestimating and overestimating the values of software effort at early phases of software development. This will lead managers to effectively bid to develop new software applications. The future study is planned to investigate the impact of locality and ensemble approaches on predicting productivity from environmental factors.

## ACKNOWLEDGMENTS

Mohammad Azzeh is grateful to the Applied Science Private University, Amman, Jordan, for the financial support granted to cover the publication fee of this research article.

Ali Bou Nassif would like to thank the University of Sharjah for supporting this research.

Cuauhtémoc López-Martín would like to thank the CUCEA, Universidad de Guadalajara, México for its support during the development of this research.